\documentclass[12pt]{article}
\usepackage{graphicx,amsmath,bm,dsfont,amssymb}

\usepackage[T2A]{fontenc}
\usepackage[cp1251]{inputenc}
\usepackage[english]{babel}
\usepackage{hyperref}
\usepackage{cite}

\usepackage{multirow}

\usepackage{subfig}
\usepackage{tabularx}

\usepackage[titletoc,title]{appendix}

\usepackage[pdftex, left=1in, right=1in, top=1in, bottom=2cm]{geometry}

\newcommand{\be}{\begin{equation}}
\newcommand{\ee}{\end{equation}}

\newcommand{\ssigma}{{\bm \sigma}}
\newcommand{\unity}{\mathds{1}}
\newcommand{\tr}{\mathrm{tr}\,}

\begin{document}

\title{Squaring parametrization of constrained and unconstrained sets of quantum states}

\author{N. Il'in$^1$\footnote{E-mail: \texttt{ilyn@mi.ras.ru}},~~ E. Shpagina$^2$,~~F. Uskov$^3$,~~O. Lychkovskiy$^{4,5,1}$}
\maketitle
\begin{center}
$^1$Steklov Mathematical Institute of Russian Academy of Sciences,\\
Gubkina str., 8, Moscow 119991, Russia
\end{center}
\begin{center}
$^2$ Bauman Moscow State Technical University,\\
2nd Baumanskaya str., 5, Moscow 105005, Russia
\end{center}
\begin{center}
$^3$ Moscow State  University, Faculty of Physics,\\
 GSP-1, 1-2 Leninskiye Gory, Moscow 119991, Russia
\end{center}
\begin{center}
$^4$ Skolkovo Institute of Science and Technology,\\
Skolkovo Innovation Center 3, Moscow  143026, Russia
\end{center}
\begin{center}
$^5$ Russian Quantum Center, \\ Novaya St. 100A, Skolkovo, Moscow
Region, 143025, Russia
\end{center}

\begin{abstract}
A mixed quantum state is represented by a Hermitian positive semi-definite operator $\rho$ with unit trace. The positivity requirement is responsible for a highly nontrivial geometry of the set of quantum states. A known way to satisfy this requirement automatically is to use the map $\rho=\tau^2/\tr\tau^2$, where $\tau$ can be an arbitrary Hermitian operator. We elaborate the parametrization of the set of quantum states induced by the parametrization of the linear space of Hermitian operators by virtue of this map. In particular, we derive an equation for the boundary of the set. Further, we discuss how this parametrization can be applied to a set of quantum states constrained by some symmetry, or, more generally, some linear condition.  As an example, we consider the parametrization of sets of Werner states of qubits.
\end{abstract}

\section{Introduction}

A quantum state of a system with a Hilbert space ${\mathcal H}={\mathbb C}^{D}$ with a finite dimension $D$ is represented by a density operator $\rho\in{\mathbb C}^{D\times D}$ which should have unit trace and be Hermitian and positive semi-definite,
\be\label{rho conditions}
\tr \rho=1,~~~\rho^\dagger=\rho,~~~ \rho\geq0.
\ee
The latter non-linear condition is responsible for an extremely complicated structure of the set ${\mathbb M}$ of quantum states which shows up for $D>2$
\cite{BengtssonZyczkowski}. To describe the shape of this set is an important problem with numerous applications in quantum information and condensed matter theory.
In particular, it is often desirable to introduce a {\it parametrization} of ${\mathbb M}$, i.e. a map from some subset of ${\mathbb R}^{D^2-1}$ to ${\mathbb M}$. If a system under consideration is a many-body system and ${\mathcal H}$ is a tensor product of one-body Hilbert spaces, one would further wish to have a parametrization with this tensor product structure built in. Unfortunately, widely used parametrizations \cite{bruning2012parametrizations} either fail to explicitly  incorporate the tensor product structure or impose the positivity condition in a rather opaque and computationally demanding form.


The purpose of the present paper is to fill this gap by elaborating a parametrization of the set  ${\mathbb M}$ of quantum states which can account for the positivity in a straightforward manner and is well-suited for many-body systems. A starting point for our reasoning is an observation made in \cite{Bengtsson} that any density operator $\rho$ can be expressed as
\be\label{main idea}
\rho=\tau^2/\tr \tau^2,
\ee
where $\tau$ is some Hermitian operator.
Obviously the r.h.s. of this equation satisfies all three conditions \eqref{rho conditions}. Eq. \eqref{main idea} establishes a map ${\mathbb H} \rightarrow {\mathbb M}$ between the real linear space ${\mathbb H}$ of Hermitian operators, and the set ${\mathbb M}$ of quantum states. This map has been used to introduce a measure in ${\mathbb M}$ induced by a measure in  ${\mathbb H}$ \cite{zyczkowski2001induced}.\footnote{To be more exact, a slightly different map of the form $\rho=A A^\dagger/\tr \left(A A^\dagger\right)$ with $A$ being an arbitrary linear operator has been used in \cite{zyczkowski2001induced}.} Here we focus on the induced parametrization rather than the induced measure. Namely, we employ the fact that  ${\mathbb H}$ is easily parametrized in a manner preserving tensor structure \cite{MahlerWeberrub}. This, in turn, induces a parametrization of ${\mathbb M}$ through the map \eqref{main idea}. We will use the term ``squaring parametrization'' for any parametrization obtained in this way.

We construct the squaring parametrization of ${\mathbb M}$, discuss its properties and its relation to the widely used Bloch vector parametrization \cite{kimura2003,byrd2003characterization}  in  Section \ref{sec:unconstrained}. In particular, within this parametrization we derive an equation for the boundary $\partial{\mathbb M}$ of ${\mathbb M}$.

In Section \ref{sec:constrained} we study the squaring parametrization of the the set of quantum states subject to linear constraints (e.g. symmetry constraints). The usage and merits of the squaring parametrization are exemplified in the case of rotationally invariant states of two and three spins $1/2$ (qubits). We conclude with the outlook of possible applications of the squaring parametrization in Section \ref{sec:discussion}. Some technical results are relegated to the Appendix.


\section{An unconstrained set of quantum states\label{sec:unconstrained}}

\subsection{Preliminaries}

We start from reminding basic facts concerning ${\mathbb M}$ and ${\mathbb H}$ \cite{BengtssonZyczkowski,MahlerWeberrub,Holevo}. The real linear space ${\mathbb H}$ of Hermitian operators acting in the Hilbert space ${\mathcal H}={\mathbb C}^{D}$ has dimension $D^2$. One can introduce a scalar product in ${\mathbb H}$ according to
\be\label{scalar product}
(\lambda, \lambda') \equiv D^{-2}\,\, \tr (\lambda \lambda'), ~~~\lambda,\, \lambda'\in\mathbb H.
\ee
${\mathbb H}$ is a real inner product space with respect to this scalar product. One can always select in ${\mathbb H}$ an orthonormal basis consisting of a identity operator, $\unity$, and $D^2-1$ generators $\lambda_i$ of the $SU(D)$ group, satisfying

\begin{align}
\lambda_i^\dagger= & \, \lambda_i,\\
\tr(\lambda_i\lambda_j)= & \, \delta_{ij} D^2, \label{tr cond} \\
\tr\lambda_i= & \, 0,  \\
[\lambda_i,\lambda_j]= & \,  2if_{ijk}\lambda_k, \label{def2}\\
\{\lambda_i,\lambda_j\}= & \,  2\delta_{ij}\unity+2d_{ijk}\lambda_k. \label{def3}
\end{align}
Here and in what follows indexes $i,j,k$ run from $1$ to  $(D^2-1)$, a summation over repeated indexes is implied, and $f_{ijk}$ and $d_{ijk}$  are totally antisymmetric and symmetric tensors, respectively. 
Relations \eqref{tr cond}, (\ref{def2}) and (\ref{def3}) can be combined,
\be\label{lambda lambda}
\lambda_i\lambda_j=\delta_{ij}\unity+if_{ijk}\lambda_k+d_{ijk}\lambda_k.
\ee

In the simplest case of a single spin $1/2$ there are three generators which can be chosen to be the Pauli matrices, $\sigma^\alpha$. In the case of a many-body system its Hilbert space ${\mathcal H}$ is a tensor product of Hilbert spaces of individual constituents, and one can choose the generators $\lambda_i$ which inherit this tensor product structure. E.g.  a system consisting of $N$ spins $1/2$ has a Hilbert space ${\mathcal H}=\left({\mathbb C}^2\right)^{\otimes N}$, $D=2^N$, and one can choose
\be
\lambda_i=\sigma_1^{\mu_{1\,i}} \otimes \sigma_2^{\mu_{2\,i}}\otimes\dots\otimes\sigma_N^{\mu_{N\,i}},
\ee
where $\sigma_n^\mu$ acts in the space of the $n$'th spin, and is equal to the $\mu$'th Pauli matrix in the case of $\mu=1,2,3$ or an identity operator in the case of $\mu=0$.  Index $i$ enumerates all possible combinations  $\{\mu_{1\,i},\mu_{2\,i},\dots,\mu_{N\,i}\}$ except one which consists of $N$ zeros.

The set ${\mathbb M}$  of all density operators $\rho$ defined according to eq. \eqref{rho conditions} is a convex set of dimension $D^2-1$ embedded in ${\mathbb H}$. The positivity condition implies that inner points of $\mathbb M$ have $D$ strictly positive eigenvalues while points on its boundary, $\mathbb M$, have at least one zero eigenvalue. The extremal points of $\mathbb M$ are pure states, i.e rank-one projectors, $\rho^2=\rho$.

\subsection{Squaring parametrization}

As is obvious from the above discussion, any quantum state $\rho$ can be expanded as
\be\label{rho}
\rho=\frac1D (\unity+a_i\lambda_i),
\ee
where $a_i,~i=1,2,\dots D^2-1$ are real parameters. Such parametrization known as the Bloch or the coherence vector parametrization \cite{kimura2003,byrd2003characterization} does not ensure the positivity automatically. The positivity condition, $\rho \geq 0$, is fulfilled if and only if the Bloch vector $a=(a_1,a_2,\dots,a_{D^2-1})$ satisfies a set of $D-1$ inequalities \cite{kimura2003,byrd2003characterization}. The first inequality of this set reads
\be
a^2\leq D-1,~~~{\rm where}~~~a^2\equiv a_i\,a_i.
\ee
It ensures that $\tr \rho^2\leq 1$. Other inequalities are defined recursively, $j$'th inequality containing a polynomial in $a_i$ of the degree $j+1$. The Bloch vector $a$ corresponds to the state on the boundary of  ${\mathbb M}$ if and only if at least one of these inequalities saturates (i.e. turns into an equality). The set of Bloch vectors subject to the aforementioned inequalities is isomorphous to the set ${\mathbb M}$ of quantum states, thus we will identify these sets.

An alternative way do describe ${\mathbb M}$ is to employ the squaring parametrization which has an advantage of imposing positivity automatically. It is defined as follows. One introduces an auxiliary Hermitian operator $\tau$ parametrized by a real vector $b=(b_1,b_2,\dots b_{D^2-1})\in {\mathbb R}^{D^2-1}$,
\be\label{tau}
\tau=\frac1D (\unity+b_i\lambda_i).
\ee
The density operator $\rho$ is then given by eq. \eqref{main idea} \cite{Bengtsson,zyczkowski2001induced}.  Obviously for any $b\in{\mathbb R}^{D^2-1}$ the conditions \eqref{rho conditions} are satisfied. Conversely, for any  density operator $\rho$ one can choose $\tau =\sqrt{\rho}/\tr\sqrt{\rho}$ (where $\sqrt{\dots}$ denotes a non-negative operator square root) and find a corresponding vector $b$. Thus we have established a surjective map ${\mathbb R}^{D^2-1}\rightarrow \mathbb M$ which constitutes the squaring parametrization. The explicit mapping between the auxiliary vector $b$ and the Bloch vector $a$ reads
\be\label{main result 1}
a_i=\frac{2b_i+d_{ijk}b_j b_k}{1+b^2},~~~{\rm where}~~~b^2\equiv b_i\,b_i.
\ee

It should be emphasized that in general the established map, which we will denote as $a(b)$, is not a one-to-one correspondence: A single Bloch vector $a$ can be obtained from several auxiliary vectors $b$ (see examples in the next Section). This is the price for the straightforward accounting for the positivity condition. However for pure states $b$ is unique and equal to $a$, which follows from $\rho^2=\rho$. The set of $D^2$ equations which determine the Bloch vector for pure states read \cite{byrd2003characterization}
\begin{align}\label{pure states equation}
a^2 = & \, D-1,\\
(D-2)a_i = & \, d_{ijk}a_j a_k.
\end{align}



As an immediate corollary of the squaring parametrization we obtain an equation satisfied by the boundary ${\partial\mathbb M}$ of the convex set  ${\mathbb M}$. A point of the boundary is a critical point of the map $a(b)$, hence the Jacobian of this map should be zero on the boundary:
\be\label{Jacobian}
{\rm det}\left|\left|\frac{\partial a_i}{\partial b_j}\right|\right|=0.
\ee
Of course some of the solutions of this equation can be inner points of ${\mathbb M}$. The usability of this equation will be exemplified in the next Section.

\section{Constrained sets of quantum states\label{sec:constrained}}

\subsection{General remarks}

Often one is interested in a set of quantum states constrained by some condition, e.g. some symmetry requirement. Alternatively, one may wish to visualize the unconstrained set $\mathbb M$ by plotting its two- or three-dimensional sections. We focus here on the most ubiquitous case of linear conditions. A linear condition can be imposed by choosing  a linear subspace ${\mathbb H}'\subset {\mathbb H}$ and defining the constrained set of quantum states ${\mathbb M}'$ as a section of $\mathbb M$ by  ${\mathbb H}'$, i.e. ${\mathbb M}'={\mathbb M}\cap{\mathbb H}'$. This way one obtains, in particular, sets of quantum states symmetric under a certain unitary transformation.

At the first sight, the machinery developed in the previous section can be directly applied in the constrained case by substituting ${\mathbb H}$ by ${\mathbb H}'$. One should be cautious, however, since the constraint can introduce important new features of the geometry of ${\mathbb M}'$. One important novel feature is that while the constrained set ${\mathbb M}'$ is still a convex set, its extreme points need not be pure states. This point will be exemplified and discussed in more detail in what follows. More importantly, a linear constraint, generally speaking, can ruin a tensor product structure of ${\mathbb H}$. Nevertheless, the squaring parametrization can be adapted for the constrained case while retaining most of its attractive features. In particular,  the boundary $\partial{\mathbb M}'$ of the constrained set ${\mathbb M}'$ will still be given by eq. \eqref{Jacobian}. We do not attempt to systematically describe a squaring parametrization for a general liner constraint in the present paper. Rather, we consider in detail an instructive example -- the parametrization of sets of Werner states of qubits.

\subsection{Set of Werner states of qubits\label{subsec: Werner}}

A Werner or rotationally invariant state of $N$ spins $1/2$ (or qubits\footnote{We use terms ``spin $1/2$'' and ``qubit'' interchangeably throughout the paper.}) is a quantum state invariant under any unitary transformation of the form $ U^{\otimes N}$, where $U$ is a unitary rotation in the space of a single spin \cite{eggeling2001separability}. The space of Werner states is rather well studied for a moderate number of spins \cite{eggeling2001separability,suzuki2012symmetric,johnson2013compatible}  or  under the additional permutation symmetry \cite{lyons2011entanglement}. We employ Werner states as a convenient playground to visualize the squaring parametrization and demonstrate its merits.

We find it convenient to expand Werner states in a basis which makes explicit their symmetry  but is not normalized. The basic building blocks of this basis are scalar and triple products of $2\times2$ sigma matrices of different spins:
\begin{align}\label{scalarvectorprod}
(\ssigma_n \ssigma_m)\equiv & \, \delta_{\alpha \beta }\, \sigma_n^\alpha \otimes \sigma_m^\beta= \sigma_n^\alpha \otimes \sigma_m^\alpha,\nonumber\\
(\ssigma_n \ssigma_m \ssigma_l)\equiv & \,  \varepsilon_{\alpha \beta \gamma}\, \sigma_n^\alpha \otimes \sigma_m^\beta \otimes \sigma_l^\gamma.
\end{align}
Here and in what follows lower indexes of the $\sigma$-matrices label qubits. The upper indexes $\alpha,\beta$ and $\gamma$ denote the $x,y,z$ components of the $\sigma$-matrices; they are always repeated which implies summation. In the reminder of the paper including Appendix we will omit the tensor product notation and substitute the identity operator $\unity$ of any dimension by $1$. In the case of two or three spins considered below in detail the basis in ${\mathbb H}'$ consists of operators of the form \eqref{scalarvectorprod} and the identity operator. For larger number of spins the basis operators involve products of operators of the form \eqref{scalarvectorprod} such as $(\ssigma_1 \ssigma_2)(\ssigma_3 \ssigma_4 \ssigma_5)$.  Some remarks regarding the case of arbitrary number of spins as well as some explicit expressions for the pairwise products of the basis operators analogue to eq. \eqref{lambda lambda} can be found in the Appendix~\ref{Appendix A}.

\subsubsection{Two qubits}

\begin{figure}
  \centering
  \includegraphics[width=0.6 \linewidth]{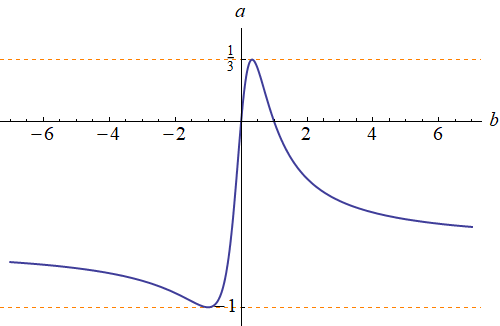}
  \caption{  \label{fig1}
Squaring parametrization for the set of Werner states of two qubits, eq. \eqref{rho two spins}. Plotted is the map $a(b)$ given by eq. \eqref{map 2 spins}. While the auxiliary parameter $b$ runs from $-\infty$ to $+\infty$, the generalized Bloch parameter $a$ automatically admits values only in the allowed range $[-1,\frac13]$.
  }
\end{figure}

We start from a very simple example of a rotationally invariant state of two qubits. This state is given by
\be\label{rho two spins}
\rho=\frac14\left(1+ a\,\, \ssigma_1\ssigma_2 \right).
\ee
The range of the only free parameter $a$ must ensure the positivity of $\rho$. To find this range we introduce the auxiliary operator $\tau$,
\be
\tau=1+ b\,\, \ssigma_1\ssigma_2
\ee
with $b\in  \mathbb{R}$ and plug it into eq. \eqref{main idea}. Note that here and in the reminder of the paper we do not impose the condition $\tr \tau=1$ implied by eq. \eqref{tau} which is of course a matter of convenience. The map $a(b)$ is found from eq. \eqref{main idea} with the use of eq. \eqref{q1} and reads
\be\label{map 2 spins}
a=\frac{2b(1-b)}{1+3b^2}.
\ee
The latter rational function has a maximum of $\frac13$ and a minimum of $(-1)$, see Fig. \ref{fig1}.
Thus eq. \eqref{rho two spins} defines a legitimate density matrix if and only if $a\in[-1,\frac13]$.

Several observations related to the previous discussion are in order.
\begin{itemize}
\item The boundary $\partial{\mathbb M}'=\{-1,\frac13\}$ of ${\mathbb M}'=[-1,\frac13]$ satisfies the equation \eqref{Jacobian} which in the present case reduces to $da/db=0$.
\item Each $a\in{\mathbb M}'$ except extreme points and the point $a=-2/3$ has two preimage points~$b$.
\item While one of the extreme points of ${\mathbb M}'$, $a=-1$, corresponds to the pure state, another one, $a=1/3$, corresponds to the mixed state. This is in contrast to the unconstrained case where all extreme points correspond to pure states. In fact, in the present case two extreme points correspond to the states with a definite total spin (0 and 1, respectively).
\end{itemize}

The latter point deserves a special remark. One can see that for a constrained space of states, $\mathcal{M}'$, the condition $\rho^2=\rho$, which have led, in particular, to eqs. \eqref{pure states equation} \cite{byrd2003characterization}, does not necessarily determine all extreme points. Instead, some of the extreme points can be described by density matrices which are equal, up to a numerical factor, to a projector with the rank $r\geq2$. This is equivalent to a condition
\be\label{projector equations generalized}
\rho^2= r^{-1}\,\rho
\ee
with an unknown integer $r$, $1\leq r<D$. This equation is sufficient to determine all extreme points in all specific cases considered in the present paper, as will be explicitly demonstrated. However, this is not the case for a general linear constraint. Furthermore, eq. \eqref{projector equations generalized} can produce additional solutions which do not correspond to extreme points, as will be the seen in examples with three qubits below. In the present case of two qubits one obtains from eq. \eqref{projector equations generalized} two extreme points,
\begin{align}
  \rho_0= & \frac14\left(1- \ssigma_1\ssigma_2 \right), & r=1, \\
  \rho_1= & \frac14\left(1+\frac13\, \ssigma_1\ssigma_2 \right), & r=3.
\end{align}
in agreement with the analysis based on the squaring parametrization.

\subsubsection{Translation-invariant Werner states of three qubits}

\begin{figure}
\center{
 \includegraphics[width=0.7 \linewidth]{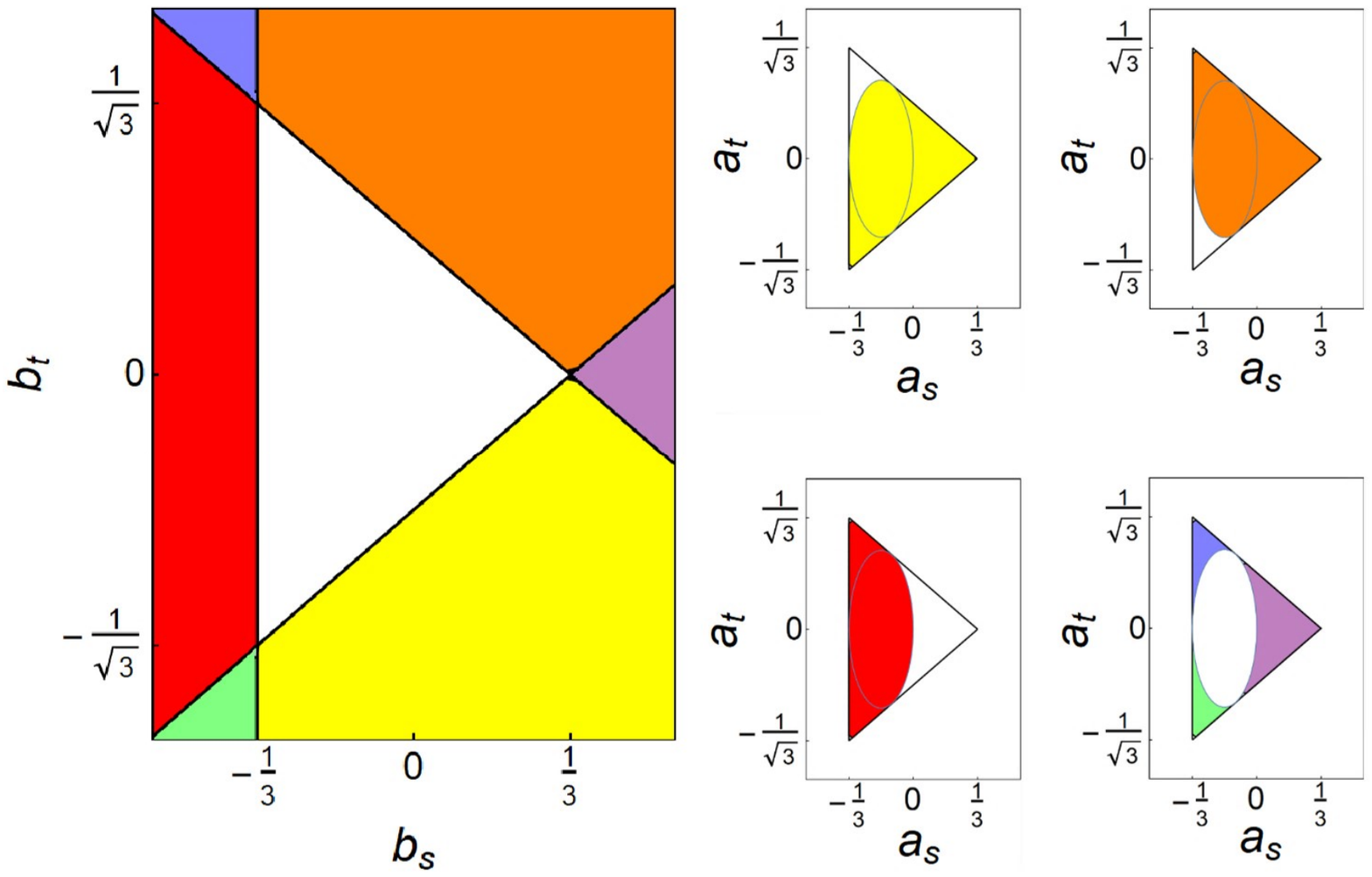}
 }
 \caption{  \label{fig2}
  Squaring parametrization for the set of translation-invariant Werner states of three qubits, eq. \eqref{rho three spins translation invariant}.
 The map $a(b)$ given by \eqref{triangle map} is shown schematically: A point of a given color in the $(b_s,b_t)$ plane is mapped to the point of the same color in the $(a_s,a_t)$ plane. Each point in the  $(a_s,a_t)$ plane has several preimage points in the $(b_s,b_t)$ plane, thus several sheets are required to visualize the map. The interior of the triangle is mapped on itself (not shown).
  }
\end{figure}

A  Werner (rotationally invariant) state of three qubits is given by
\be\label{rho three spins}
\rho=\frac18\left(1+ a_{12}\, \ssigma_1\ssigma_2+ a_{23}\, \ssigma_2\ssigma_3+ a_{31}\, \ssigma_3\ssigma_1+a_{123}\,\ssigma_1\ssigma_2\ssigma_3\right).
\ee
We would like to reduce the number of parameters from four to three or two for the purpose of visualization. To this end we impose additional symmetries. In the present subsection we require states to be translational invariant which leads to a two-dimensional constrained  set of quantum states ${\mathbb M}'$. In the next subsection we impose $T$-invariance and get a three-dimensional ${\mathbb M}'$.

A translational symmetry is a symmetry  under  cyclic permutations of qubits, $(1,2,3)\rightarrow(3,1,2) \rightarrow(2,3,1)$. Imposing  this symmetry, we get a two-dimensional set of states of the form
\be\label{rho three spins translation invariant}
\rho=\frac18\left(1+ a_s(\ssigma_1\ssigma_2+  \ssigma_2\ssigma_3+  \ssigma_3\ssigma_1)+a_t\,\ssigma_1\ssigma_2\ssigma_3\right).
\ee
Introducing
\be
\tau=1+ b_s(\ssigma_1\ssigma_2+  \ssigma_2\ssigma_3+  \ssigma_3\ssigma_1)+b_t\,\ssigma_1\ssigma_2\ssigma_3,
\ee
we obtain from eq. \eqref{main idea} with the use of eqs. \eqref{q1}--\eqref{q5a}, \eqref{q3}
\begin{align}
a_s&=2\,\frac{b_s-b_t^2}{1+9 b_s^2+6 b_t^2}, \nonumber \\
a_t &=2 \frac{b_t(1-3b_s)}{1+9 b_s^2+6 b_t^2}.\label{triangle map}
\end{align}
These equations determine the shape of ${\mathbb M}'$ which  is a triangle shown in Fig. \ref{fig2}. A direct way to see this is to find the boundary  $\partial{\mathbb M}'$ from eq. \eqref{Jacobian}
which in the present case reads
\be\label{triangle boundary}
\det \left|\left|\frac {\partial (a_s,a_t)}{\partial (b_s,b_t)}\right|\right| =
		4\frac{(1+3 b_s)((1-3b_s)^2-12 b_t^2)}{(1+9 b_s^2+6 b_t^2)^3}=0.
\ee
The solutions of this equation determine three lines in the space of $b$ variables:
\begin{align}
b_s&=-\frac13,\nonumber\\
b_t&=\frac1{ 2 \sqrt3}(1-3b_s),\nonumber\\
b_t&=-\frac1{ 2 \sqrt3}(1-3b_s).\label{trinagle b coordinates}	
\end{align}
The map \eqref{triangle map} converts these lines to three line segments,
	\begin{align}
		a_s &=-\frac13,& a_t & \in[-\frac1{\sqrt{3}},\frac1{\sqrt{3}}], \nonumber \\
		a_t &=\frac1{ 2\sqrt3} (1-3a_s), & a_s & \in[-\frac13,\frac13],\nonumber \\
		a_t &=-\frac1{ 2\sqrt3} (1-3a_s), & a_s & \in[-\frac13,\frac13].\label{trinagle a coordinates}
	\end{align}
The map \eqref{triangle map} is illustrated in fig. \eqref{fig2}.


Remarkably, eqs. \eqref{trinagle a coordinates} replicate eqs. \eqref{trinagle b coordinates}, up to the allowed range of variables. The extreme points of ${\mathbb M}'$ are the vertexes of the triangle. An alternative way to determine them is to use eq. \eqref{projector equations generalized} as shown in Appendix \ref{Appendix B}.




\subsubsection{$T$-invariant Werner states of three qubits}

\begin{figure}[t]
  \centering
  \includegraphics[width=0.7 \linewidth]{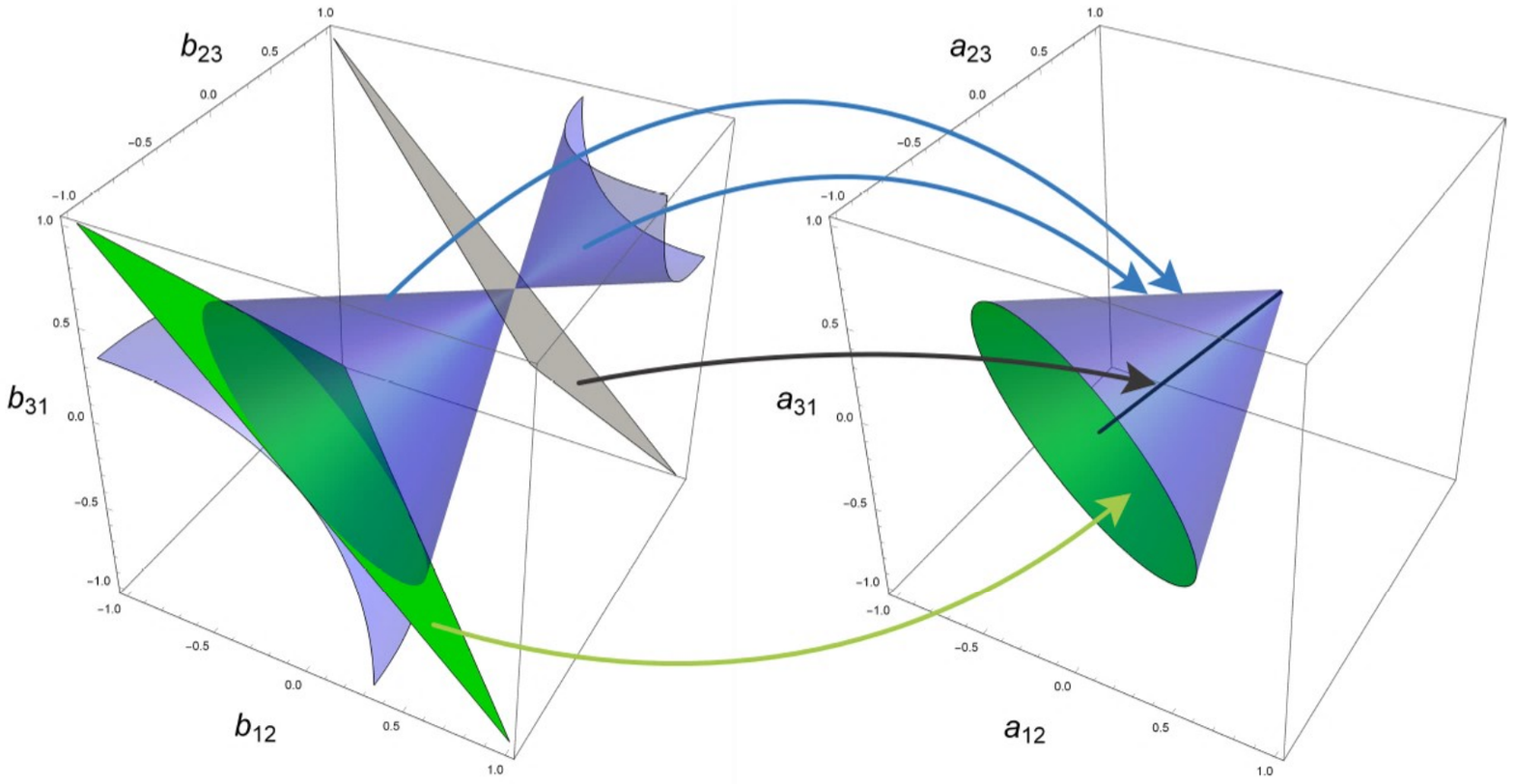}
 \caption{  \label{fig3}
  Squaring parametrization for the set of $T$-invariant Werner states of three qubits, eq. \eqref{rho three spins T-invariant}.
In the left panel the solutions of the equation  \eqref{Jacobian} for the boundary  $\partial{\mathbb M}'$ are shown in the space of auxiliary parameters $(b_{12},b_{23},b_{31})$. The same solutions in the space of generalized Bloch vectors $(a_{12},a_{23},a_{31})$ are shown in the right panel. The set ${\mathbb M}'$ is a truncated cone.  Arrows illustrate  the mapping \eqref{parametrization three spins} applied to solutions of eq. \eqref{Jacobian}.
  }
\end{figure}

Now we turn to the case of Werner states of three qubits invariant under time reversal. This transformation acts on products of sigma matrices as follow:
$T(\ssigma_i\ssigma_k)=\ssigma_i\ssigma_k$ and $T(\ssigma_1\ssigma_2\ssigma_3)=-\ssigma_1\ssigma_2\ssigma_3$. Hence, the condition $T(\rho)=\rho$ with $\rho$ given by eq. \eqref{rho three spins} implies $a_{123}=0$,  and we obtain a three-dimensional set of states of the form
\be\label{rho three spins T-invariant}
\rho=\frac18\left( 1+ a_{12}\, \ssigma_1\ssigma_2+ a_{23}\, \ssigma_2\ssigma_3+ a_{13}\, \ssigma_1\ssigma_3\right).
\ee
Introducing
\be
\tau=1+ b_{12}\, \ssigma_1\ssigma_2+ b_{23}\, \ssigma_2\ssigma_3+ b_{13}\, \ssigma_1\ssigma_3
\ee
we get
\be\label{parametrization three spins}
a_{12}=2\,\frac{b_{12}-b_{12}^2+b_{23}b_{13}}{1+3(b_{12}^2+b_{23}^2+b_{13}^2)},
\ee
and analogous formulae for $a_{23}$ and $a_{13}$.

To determine the boundary  $\partial{\mathbb M}'$ we use  eq. \eqref{Jacobian} which in this case reads
\be\label{zeroJ}
\det \left|\left|\frac {\partial (a_{12},a_{23},a_{13})}{\partial (b_{12},b_{23},b_{13})}\right|\right| = 24 \frac{(A-1)(A+1)(B+ \frac23 A - C- \frac13)}{(3 B+1)^4}=0,
\ee
where
\begin{align}
A&=b_{12}+b_{23}+b_{13},\\
B&=b_{12}^2+b_{23}^2+b_{13}^2,\\
C&= 2( b_{12}b_{23}+b_{23}b_{13}+b_{13}b_{12}).
\end{align}
Solutions of  eq. \eqref{zeroJ}  have the following form:
\begin{align}
	b_{12}^2+b_{23}^2+b_{13}^2-2( b_{12}b_{23}+b_{23}b_{13}+b_{13}b_{12})+ \frac23(b_{12}+b_{23}+b_{13})&=\frac13,\label{cone1} \\
	b_{12}+b_{13}+b_{23}&=-1, \label{plane2}\\
	b_{12}+b_{13}+b_{23}&=1. \label{plane1}
\end{align}
Eq.  \eqref{cone1} describes a double cone with a vertex with coordinates
$
b_{12}=b_{13}=b_{23}=\frac13
$,
while eqs. \eqref{plane2}, \eqref{plane1} -- two parallel planes.
The map \eqref{parametrization three spins} converts the double cone to a truncated cone, one of the planes to the base of this truncated cone and another plane to the altitude of this cone, as shown in Fig. \ref{fig3}. In contrast to the previously considered cases, some of the solutions of eq. \eqref{Jacobian} correspond to inner points of ${\mathcal M}'$. Finding extreme points of ${\mathcal M}'$ with the help of eq. \eqref{projector equations generalized} is described in the Appendix \ref{Appendix C}.

\section{Summary and outlook\label{sec:discussion}}

To summarize, the squaring parametrization is a surjective nonlinear map from ${\mathbb R}^{D^2-1}$ to a set ${\mathbb M}$ of (in general, mixed) quantum states. This map is explicitly given by eq. \eqref{main result 1} which maps each point of ${\mathbb R}^{D^2-1}$  to a legitimate Bloch vector. The squaring parametrization has several attractive features. First, it automatically accounts for the positivity of a density operator. Second, it produces as a byproduct an equation for boundary of ${\mathbb M}$, see eq. \eqref{Jacobian}. Third, a tensor product structure of a many-body system can be explicitly retained within this parametrization. Finally, the squaring parametrization can be adapted to describe sets of quantum states constrained by a certain symmetry or, more general, a linear constraint.

We believe that the squaring parametrization can be useful in a wide range of problems where the quantum density matrix of a sufficiently large dimension plays a role. This includes questions of separability, entanglement and quantum information processing (especially with higher-dimensional qudits \cite{kiktenko2015multilevel} instead of qubits). A special mention in this context is deserved by a variational technique which is based on variation of  a reduced density matrix (instead of a many-body wave function) and bounds the ground state energy from below (not from above) \cite{Maziotti}. The squaring parametrization seems to be especially well-suited for this technique.

\vspace{1.5 em}

{\it Acknowledgements.}~~ The support from the Russian Science Foundation under the grant N$^{\rm o}$ 17-11-01388 is acknowledged.

\begin{appendices}

\section{Rotationally invariant operators in the space of qubits\label{Appendix A}}

In order to use the squaring parametrization for Werner states one needs to be able to convolute  products of scalar and triple products of sigma matrices. Here we give a list of such convolutions involving at most four qubits.

\begin{align}
  (\ssigma_1\ssigma_2)^2 = & \,\,\,  3\
  -2 (\ssigma_1\ssigma_2)				\label{q1}\\
  %
  (\ssigma_1\ssigma_2)(\ssigma_2\ssigma_3) = &\
  -i(\ssigma_1\ssigma_2\ssigma_3)\
  +(\ssigma_1\ssigma_3)					\label{q2}\\
  %
  (\ssigma_1\ssigma_2)(\ssigma_1\ssigma_2\ssigma_3) = &\
  -(\ssigma_1\ssigma_2\ssigma_3)\
  -2i(\ssigma_1\ssigma_3)\
  +2i(\ssigma_2\ssigma_3)				\label{q5}\\
  %
  (\ssigma_1\ssigma_2\ssigma_3)(\ssigma_1\ssigma_2) = &\
  -(\ssigma_1\ssigma_2\ssigma_3)\
  +2i(\ssigma_1\ssigma_3)\
  -2i(\ssigma_2\ssigma_3)				\label{q5a}\\
  %
  (\ssigma_1\ssigma_2)(\ssigma_2\ssigma_3\ssigma_4) = & \
  (\ssigma_1\ssigma_3\ssigma_4)\
  -i(\ssigma_1\ssigma_3)(\ssigma_2\ssigma_4)\
  +i(\ssigma_1\ssigma_4)(\ssigma_2\ssigma_3)		\label{q4}\\
  %
  (\ssigma_2\ssigma_3\ssigma_4)(\ssigma_1\ssigma_2) =  &\
  (\ssigma_1\ssigma_3\ssigma_4)\
  +i(\ssigma_1\ssigma_3)(\ssigma_2\ssigma_4)\
  -i(\ssigma_1\ssigma_4)(\ssigma_2\ssigma_3)		\label{q4a}\\
  %
  (\ssigma_1\ssigma_2\ssigma_3)^2 = &\,\,\,6\
  -2(\ssigma_1\ssigma_2)\
  -2(\ssigma_1\ssigma_3)\
  -2(\ssigma_2\ssigma_3)				\label{q3}\\
  %
  (\ssigma_1\ssigma_2\ssigma_3)(\ssigma_1\ssigma_2\ssigma_4) = &\
  +i(\ssigma_1\ssigma_3\ssigma_4)\
  +i(\ssigma_2\ssigma_3\ssigma_4) 		\nonumber\\  & \
  -(\ssigma_1\ssigma_3)(\ssigma_2\ssigma_4) \
  -(\ssigma_1\ssigma_4)(\ssigma_2\ssigma_3)
  2 (\ssigma_3\ssigma_4)				\label{q6}\\
  %
  (\ssigma_1\ssigma_2\ssigma_3)(\ssigma_1\ssigma_4\ssigma_5) = & \
  -i(\ssigma_1 \ssigma_2)(\ssigma_3 \ssigma_4 \ssigma_5) \
  +i(\ssigma_1 \ssigma_3 )(\ssigma_2 \ssigma_4 \ssigma_5) \nonumber\\  & \
  +(\ssigma_2 \ssigma_4)(\ssigma_3 \ssigma_5)\
  -(\ssigma_2 \ssigma_5)(\ssigma_3 \ssigma_4) 		\label{q7}\\
  %
  (\ssigma_1\ssigma_2)(\ssigma_2\ssigma_3)(\ssigma_3\ssigma_4) =& \
  -i(\ssigma_1\ssigma_2\ssigma_4)\
  -i(\ssigma_1\ssigma_3\ssigma_4)\
  (\ssigma_1\ssigma_4)		\nonumber\\&\
  +(\ssigma_1\ssigma_4)(\ssigma_2\ssigma_3)\
  -(\ssigma_1\ssigma_3)(\ssigma_2\ssigma_4)		\label{q8}\\
  %
  (\ssigma_1\ssigma_2)(\ssigma_3\ssigma_4)(\ssigma_1\ssigma_3)(\ssigma_2\ssigma_4) = & \,\,\,3\
  +i(\ssigma_1\ssigma_2\ssigma_3)\
  -i(\ssigma_1\ssigma_2\ssigma_4)\
  +i(\ssigma_1\ssigma_3\ssigma_4)\
  -i(\ssigma_2\ssigma_3\ssigma_4)	\nonumber\\&\
  -2(\ssigma_1\ssigma_2)\
  -2(\ssigma_1\ssigma_3)\
  +2(\ssigma_1\ssigma_4)		\nonumber\\&\
  +2(\ssigma_2\ssigma_3)\
  -2(\ssigma_2\ssigma_4)\
  -2(\ssigma_3\ssigma_4)		\nonumber\\&\
  +(\ssigma_1\ssigma_2)(\ssigma_3\ssigma_4)\
  +(\ssigma_1\ssigma_3)(\ssigma_2\ssigma_4) 		\label{q9}
\end{align}

It should be noted that the first nine of these equalities suffice to evaluate all other combinations of scalar and triple products, in particular, eqs. \eqref{q8},\eqref{q9} iteratively without exploiting the properties of $\sigma$-matrices.

Also observe that eq. \eqref{q9} implies that operators $ (\ssigma_1\ssigma_2)(\ssigma_3\ssigma_4)$ and $(\ssigma_1\ssigma_3)(\ssigma_2\ssigma_4)$ are not orthogonal with respect to the scalar product \eqref{scalar product}. This implies that in the case of the number of qubits higher than three on has either to employ an additional orthogonalization procedure or cope with a nonorthogonal basis.

\section{Extreme points for the translation invariant Werner states of three qubits\label{Appendix B}}

\begin{figure}[t]
\center{
\includegraphics[width=0.5 \linewidth]{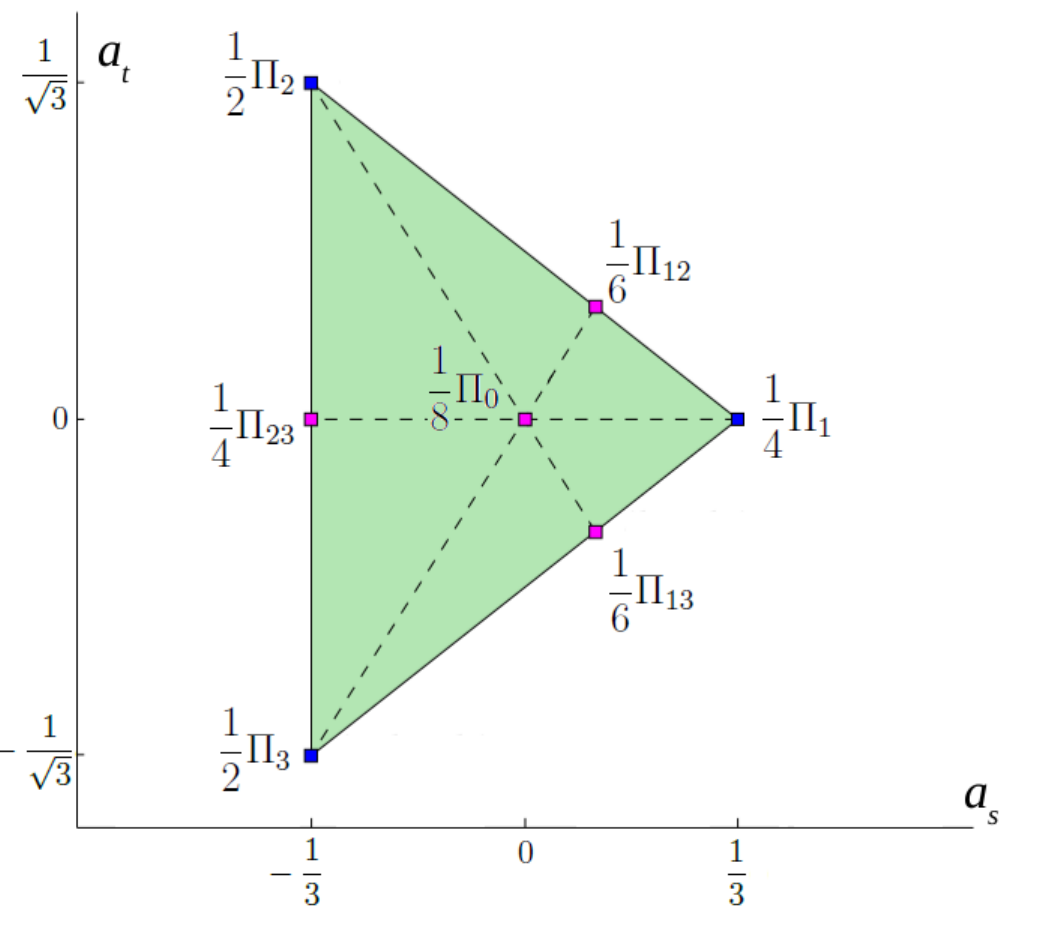}
 \caption{  \label{fig4}
   Solutions of eq. \eqref{projector equations generalized} in the case of translation-invariant Werner states of three qubits, eq. \eqref{rho three spins translation invariant}. A projectors is turned into a density matrix by normalizing it by the numerical factor $1/r$, where $r$ is the rank of the projector. Three of the solutions correspond to the extreme points of ${\mathcal M}'$ (vertexes of the triangle), others can be obtained as equally weighted linear combinations of two or three ``extreme'' solutions.
  }
  }
\end{figure}

Here we solve equation \eqref{projector equations generalized} and this way find the extreme points of the set ${\mathcal M}'$ of the  states of the form \eqref{rho three spins translation invariant}. We choose to work with projectors $\Pi$ which are related to the solutions of eq. \eqref{projector equations generalized} as $\Pi=r \rho$ and, obviously, satisfy the equation
\be\label{projector squared}
\Pi^2=\Pi.
\ee
We further introduce $c\equiv r/D$. Substituting
\be
\Pi=c(1+ a_s(\ssigma_1\ssigma_2+  \ssigma_2\ssigma_3+  \ssigma_3\ssigma_1)+a_t\,\ssigma_1\ssigma_2\ssigma_3).\label{3spinsProj}
\ee
into eq. \eqref{projector squared} and using eqs. \eqref{q1}--\eqref{q5} one obtains
\begin{align}
  c&=(1+9a_s^2+6a_t^2)^{-1}\label{eq00},\\
  a_s&=2\,c\,(a_s-a_t^2)\label{eq01},\\
  a_t&=2\,c\,a_t\,(1-3a_s)\label{eq02}.
\end{align}
The solutions of this system of equations correspond to seven  projectors:
\begin{align}
  \Pi_0&=1\label{Pi0triangle}\\
  \Pi_1&=\frac{1}{2}\left(1+\frac{1}{3}(\ssigma_1\ssigma_2+ \ssigma_2\ssigma_3+ \ssigma_3\ssigma_1)\right),\\
  \Pi_2&=\frac{1}{4}\left(1-\frac{1}{3}(\ssigma_1\ssigma_2+ \ssigma_2\ssigma_3+ \ssigma_3\ssigma_1)+\frac{1}{\sqrt{3}}\ssigma_1\ssigma_2\ssigma_3\right),\\
  \Pi_3&=\frac{1}{4}\left(1-\frac{1}{3}(\ssigma_1\ssigma_2+ \ssigma_2\ssigma_3+ \ssigma_3\ssigma_1)-\frac{1}{\sqrt{3}}\ssigma_1\ssigma_2\ssigma_3\right),\\
  \Pi_{12}&=\frac{3}{4}\left(1+\frac{1}{9}(\ssigma_1\ssigma_2+ \ssigma_2\ssigma_3+ \ssigma_3\ssigma_1)+\frac{1}{3\sqrt{3}}\ssigma_1\ssigma_2\ssigma_3\right),\\
  \Pi_{23}&=\frac{1}{2}\left(1-\frac{1}{3}(\ssigma_1\ssigma_2+ \ssigma_2\ssigma_3+ \ssigma_3\ssigma_1)\right),\\
  \Pi_{13}&=\frac{3}{4}\left(1+\frac{1}{9}(\ssigma_1\ssigma_2+ \ssigma_2\ssigma_3+ \ssigma_3\ssigma_1)-\frac{1}{3\sqrt{3}}\ssigma_1\ssigma_2\ssigma_3\right).\label{Pi13triangle}
\end{align}

Extreme points of ${\mathcal M}'$ are given by $\Pi_1$, $\Pi_2$ and $\Pi_3$. All other projectors can be represented as equally weighted  linear combinations of two or three of these ``extreme'' projectors, e.g.    $\Pi_{23}=\Pi_1+\Pi_2$. Density matrices obtained from projectors \eqref{Pi0triangle}-\eqref{Pi13triangle} are shown in Fig. \ref{fig4}.

\section{Extreme points for the $T$-invariant Werner states of three qubits\label{Appendix C}}

\begin{figure}[t]
\center{
\includegraphics[width=0.5 \linewidth]{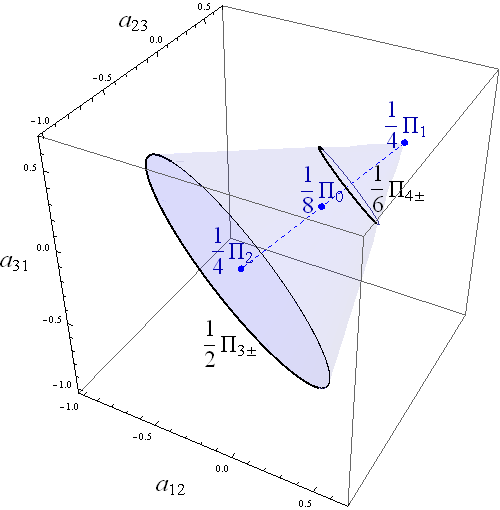}
 \caption{  \label{fig5}
   Solutions of eq. \eqref{projector equations generalized} in the case of $T$-invariant Werner states of three qubits, eq. \eqref{rho three spins T-invariant}. A projectors is turned into a density matrix by normalizing it by the numerical factor $1/r$, where $r$ is the rank of the projector. $\frac14 \Pi_1$ and $\frac12 \Pi_{3\pm}$ correspond to extreme points of ${\mathcal M}'$ (vortex and directrix, respectively).
  }
  }
\end{figure}

Here we repeat the procedure described in Appendix \ref{Appendix B} for the states of the form \eqref{rho three spins T-invariant}. We consider a projector of the form
\be
\Pi=c\,\left(1+ a_{12}\, \ssigma_1\ssigma_2+ a_{23}\, \ssigma_2\ssigma_3+ a_{31}\, \ssigma_3\ssigma_1\right)
\ee
and plug it to eq. \eqref{projector squared}. Using eqs. \eqref{q1}, \eqref{q2} we obtain equations
\begin{align}
  c & =\frac1{1+3(a_{12}^2+ a_{23}^2+a_{31}^2)},\label{c}\\
  a_{12} & =2\, \frac{a_{12}-a_{12}^2+a_{23} a_{31}}{1+3(a_{12}^2+ a_{23}^2+a_{31}^2)}\label{projector equations}
\end{align}
and two more equations which can be obtained from eq. \eqref{projector equations} by cyclic permutation of indexes in $a_{ij}$.
The solutions of these equations correspond to the following projectors:
\begin{align}
\Pi_0&=1,\label{Pi0} \\
\Pi_1&=\frac12 \left(1+ \frac13  \ssigma_1\ssigma_2+ \frac13 \, \ssigma_2\ssigma_3+ \frac13 \, \ssigma_3\ssigma_1\right), \\
\Pi_2&=\frac12 \left(1- \frac13  \ssigma_1\ssigma_2- \frac13 \, \ssigma_2\ssigma_3- \frac13 \, \ssigma_3\ssigma_1\right), \\
\Pi_{3\pm}&=\frac14 \left(1+  a_{12} \ssigma_1\ssigma_2- \frac{a_{12}\mp k_{12}+1}2 \, \ssigma_1\ssigma_3- \frac{a_{12}\pm k_{12}+1}2 \, \ssigma_2\ssigma_3\right),\\
\nonumber&k_{12} \equiv \sqrt{-3a_{12}^2-2a_{12}+1},~~~a_{12}\in[-1,\frac13],\\
\Pi_{4\pm}&=\frac34 \left(1+  a_{12} \ssigma_1\ssigma_2+ \frac{-3a_{12}\mp l_{12}+1}6 \, \ssigma_1\ssigma_3+ \frac{-3a_{12}\pm l_{12}+1}6 \, \ssigma_2\ssigma_3\right),\label{Pi4} \\ 	
\nonumber &l_{12} \equiv \sqrt{-27a_{12}^2+6a_{12}+1},~~~a_{12}\in[-\frac19,\frac13].
\end{align}
Note that $\Pi_{3\pm}$ and $\Pi_{4\pm}$ are one-parametric families of projectors parametrized by $a_{12}$.

Extreme points of ${\mathcal M}'$ are given by $\Pi_1$ (vertex) and  $\Pi_{3\pm}$ (directrix). Density matrices obtained from projectors \eqref{Pi0}-\eqref{Pi4} are shown in Fig. \ref{fig5}.

\end{appendices}

\bibliography{dmp}{}
\bibliographystyle{unsrt}
\end{document}